\documentclass[aps,preprint,10pt,twocolumn,tightenlines,showpacs]{revtex4}
\usepackage{amssymb}
\usepackage{amsfonts}
\usepackage{amsmath}
\usepackage{graphicx}

\input{tcilatex}
\begin{document}

\preprint{}
\title[ ]{Hybrid matter-wave-microwave solitons produced by the local-field
effect}
\author{Jieli Qin, Guangjiong Dong}
\affiliation{State Key Laboratory of Precision Spectroscopy, Department of Physics, East
China Normal University, Shanghai, China}
\author{Boris A. Malomed}
\affiliation{Department of Physical Electronics, School of Electrical Engineering,
Faculty of Engineering, Tel Aviv University, Ramat Aviv 69978, Israel}

\begin{abstract}
It was recently found that the electric local-field effect (LFE) can lead to
strong coupling of atomic Bose-Einstein condensates (BECs) to off-resonant
optical fields. We demonstrate that the magnetic LFE gives rise to a
previously unexplored mechanism for coupling a (pseudo)spinor BEC or fermion
gas to microwaves (MWs). We present a theory for the magnetic LFE, and find
that\ it gives rise to a short-range attractive interaction between two
components of the (pseudo) spinor, and a long-range interaction between
them. The latter interaction, resulting from deformation of the magnetic
field, is locally repulsive but globally attractive, in sharp contrast with
its counterpart for the optical LFE, produced by phase modulation of the
electric field. Our analytical results, confirmed by the numerical
computations, show that the long-range interaction gives rise to
modulational instability of the spatially uniform state, and creates stable
ground states in the form of hybrid matter-wave-microwave solitons (which
seem like one-dimensional magnetic monopoles), with a size much smaller than
the MW wavelength, even in the presence of arbitrarily strong contact
inter-component repulsion. The setting is somewhat similar to
exciton-polaritonic condensates in semiconductor microcavities. The release
of matter waves from the soliton may be used for the realization of an atom
laser. The analysis also applies to molecular BECs with rotational states
coupled by the electric MW field.
\end{abstract}

\volumeyear{year}
\volumenumber{number}
\issuenumber{number}
\eid{identifier}
\date[Date text]{date}
\received[Received text]{date}
\revised[Revised text]{date}
\accepted[Accepted text]{date}
\published[Published text]{date }
\maketitle

Ultracold atomic gases %\cite{6}
are used in a various areas, including quantum metrology and interferometry 
\cite{pri0}-\cite{pri2}, and the emulation of nonequilibrium quantum
dynamics \cite{1} and condensed-matter physics \cite{add1}-\cite{3}. They
have also drawn much interest as tunable media for quantum optics. In this
vein, co-manipulation of quantum light and matter waves has been studied in
cavities loaded with atomic Bose-Einstein condensates (BECs) \cite{ig,cavity}%
. Raman superradiance in ultracold gases trapped in a cavity was used to
generate stationary lasing with a bandwidth $<1$ MHz, and with the average
cavity photon number $<1$ \cite{8}. A mirrorless parametric resonance has
been demonstrated for atomic BEC loaded into an optical lattice (OL) \cite{m}%
. Optomechanics-induced large-scale structuring of ultracold atomic gases
has been reported in Ref. \cite{st}. The resonant interaction of laser
fields with BEC was also proposed for generating ``photonic bubbles"
emulating cosmology settings \cite{bub}.

An important feature of the interaction of light with ultracold gases is the
local-field effect (LFE), i.e., a feedback of the BEC on the light
propagation. Strong LFE can be induced in cold-atom experiments, as recently
demonstrated with the help of OLs \cite{li,d1,d2}. Usually, OLs are sturdy
structures, maintaining perfect interference fringes. However, asymmetric
matter-wave diffraction on an OL formed by counterpropagating optical fields
with unequal intensities \cite{li} could be explained only by taking into
regard deformation of the OL by the LFE \cite{d1}. %, which is critically
%important for high-precision quantum measurements \cite{can}.
Conventional rigid OLs and their deformable counterparts may be categorized
as \textquotedblleft stiff" and \textquotedblleft soft" ones. Polaritonic
solitons, produced by hybridization of coupled atomic and optical waves,
have been predicted in soft OLs \cite{d2}, being promising for the
matter-wave interferometry due to their high density \cite{si}. These
results demonstrate the potential of the soft OLs in studies of systems
combining quantum matter and photons, akin to exciton-polaritons in
microcavities \cite{excit}. Furthermore, BECs built of up to $10^{8}$ atoms
are now available \cite{107}. For such massive BECs, the refraction-index
change through the perturbation of the atomic density may be significant,
even for the laser-frequency detuning from the resonance $\gg 1$ GHz,
allowing the LFE to generate hybrid matter-wave-photonic states \cite{d2}.

The use of spinor gases opens ways for the emulation of the spin-orbit
coupling \cite{3} and quantum magnetism \cite{qmag}, as well as for the
realization of quantum matter-wave \cite{qmo} and microwave (MW) \cite{qo}
optics. In this context, the MW magnetic field is used for manipulating spin
states. Coupling different hyperfine atomic states by MWs was studied in
other contexts too, including dressed states \cite{dressed}, domain walls 
\cite{domain-wall}, %, symmetry breaking \cite{gap-sol}, periodically
%modulated coupling \cite{Niederberger},
and instabilities \cite{crossing-avoidance}. 
%\cite{crossing-avoidance}, multi-body interactions \cite{many-body}, etc.

However, manifestations of the magnetic LFE (MLFE) in quantum gases were not
studied yet, unlike its electric counterpart. In this work, we develop the
theory of the MLFE for a MW field coupled to the pseudospinor BEC. The MW
wavelength ($\gtrsim $ several mm) exceeds the typical size of the BEC by
orders of magnitude. In this situation, the BEC was considered before as a
thin slice that affects the phase of the MW field. We find that the MLFE
causes subwavelength deformations of the MW amplitude profile too, inducing
a long-range interaction between components of the pseudospinor BEC. Unlike
the electric LFE \cite{d2}, where nonlocal interaction is induced by phase
perturbations, the long-range interaction generated by the MLFE is locally
repulsive but globally attractive. The same effect leads to local attraction
between the components of the BEC, which may compete with collisional
repulsion between them. We demonstrate that these interactions create
self-trapped ground states (GSs) in the form of hybrid matter-wave-microwave
solitons, whose field component seems like that of a magnetic monopole.
Actually, the solitons realize a dissipation- and pump-free counterpart of
hybridized exciton-polariton complexes in dissipative microcavities, pumped
by external laser fields. Opposite to our case, the size of those complexes
is much larger than the polaritonic wavelength, while the effective mass of
the excitons is usually considered infinite \cite{excit}. Note that direct
dissipation-free emulation of the exciton-polariton setting is possible too
in a dual-core optical system \cite{Padova}

We consider the magnetic coupling of the MW radiation with frequency $\omega
_{L}$ to two hyperfine atomic states $\left\vert \downarrow \right\rangle $
and $\left\vert \uparrow \right\rangle $ which compose the BEC\ pseudospinor 
\cite{spinor,3}, with free Hamiltonian $\mathrm{H}_{0}=\hat{p}%
^{2}/(2m)-\left( \hbar \delta /2\right) \sigma _{3},$ where $\hbar \delta $
is the energy difference between the two states, $\sigma _{3}$ is the Pauli
matrix, and $\hat{p}$ the atomic momentum. The magnetic interaction is
governed by term 
\begin{equation}
\mathrm{H}_{\mathrm{int}}=-\left( 
\begin{matrix}
\mathbf{m}_{\mathbf{\downarrow \downarrow }} & \mathbf{m}_{\mathbf{%
\downarrow }\mathbf{\uparrow }} \\ 
\mathbf{m}_{\mathbf{\uparrow }\mathbf{\downarrow }} & \mathbf{m}_{\mathbf{%
\uparrow \uparrow }}%
\end{matrix}%
\right) \cdot \left( \mathbf{B}e^{-i\omega _{L}t}+\mathbf{B}^{\ast
}e^{i\omega _{L}t}\right) .  \label{int}
\end{equation}%
Here, $\mathbf{m}_{\mathbf{\downarrow },\mathbf{\uparrow }}$ are matrix
elements of the magnetic momentum, and the magnetic induction is $\mathbf{B}%
=\mu _{0}\mathbf{H}+\mathbf{M}$ with $\mathbf{M}=\mathbf{m}_{\mathbf{%
\downarrow }\mathbf{\uparrow }}\psi _{\mathbf{\downarrow }}^{\ast }\psi _{%
\mathbf{\uparrow }}$, with $\psi _{\mathbf{\downarrow }}$ and $\psi _{%
\mathbf{\uparrow }}$ building the pseudospinor wave function, $\left\vert
\Psi \right\rangle =\left( \psi _{\mathbf{\downarrow }}\exp \left( i\omega
_{L}t/2\right) ,\psi _{\mathbf{\uparrow }}\exp \left( -i\omega
_{L}t/2\right) \right) ^{T}$. In the rotating-wave approximation, $%
\left\vert \psi \right\rangle =\left( \psi _{\mathbf{\downarrow }},\psi _{%
\mathbf{\uparrow }}\right) ^{T}$ satisfies the system of coupled
Gross-Pitaevskii equations (GPEs):%
\begin{align}
i\hbar \frac{\partial \left\vert \psi \right\rangle }{\partial t}& =\left[ -%
\frac{\hbar ^{2}}{2m}\nabla ^{2}+\frac{\hbar \Delta }{2}\sigma _{3}-\mu
_{0}\left( 
\begin{matrix}
0 & \mathbf{m}_{\mathbf{\downarrow }\mathbf{\uparrow }}\cdot \mathbf{H}%
^{\ast } \\ 
\mathbf{m}_{\mathbf{\uparrow }\mathbf{\downarrow }}\cdot \mathbf{H} & 0%
\end{matrix}%
\right) \right.  \notag \\
& \left. -\mathbf{m}_{\mathbf{\uparrow }\mathbf{\downarrow }}\cdot \mathbf{m}%
_{\mathbf{\downarrow }\mathbf{\uparrow }}\left( 
\begin{matrix}
\left\vert \psi _{\mathbf{\uparrow }}\right\vert ^{2} & 0 \\ 
0 & \left\vert \psi _{\mathbf{\downarrow }}\right\vert ^{2}%
\end{matrix}%
\right) \right] \left\vert \psi \right\rangle ,  \label{psia}
\end{align}%
with the MW detuning from the atomic transition $\Delta =\omega _{L}-\delta $%
. Neglecting the time derivatives of $\mathbf{H}$ and $\mathbf{M}$ for the
low-frequency MW, the wave equation for $\mathbf{H}$ reduces to the
Helmholtz form, $\nabla ^{2}\mathbf{H}+\omega _{L}^{2}/c^{2}\mathbf{H}%
=-\varepsilon _{0}\omega _{L}^{2}\mathbf{M}$.

We consider a cigar-shaped condensate with effective cross-section area $S$,
subject, as usual, to tight transverse confinement, and is irradiated by two
linearly-polarized counterpropagating microwaves along the cigar's axis%
\textbf{\ }$x$. Eliminating the transverse variation of the fields under
these conditions \cite{Luca}, we reduce the coupled GPEs and Helmholtz
equation to the normalized form, 
\begin{equation}
i\frac{\partial \left\vert \phi \right\rangle }{\partial \tau }=\left[ -%
\frac{1}{2}\frac{\partial ^{2}}{\partial x^{2}}+\eta \sigma _{3}-\left( 
\begin{matrix}
\beta \left\vert \phi _{\mathbf{\uparrow }}\right\vert ^{2} & \mathcal{H}%
^{\ast } \\ 
\mathcal{H} & \beta \left\vert \phi _{\mathbf{\downarrow }}\right\vert ^{2}%
\end{matrix}%
\right) \right] \left\vert \phi \right\rangle ,  \label{phi}
\end{equation}%
\begin{equation}
\partial _{x}^{2}\mathcal{H}+\kappa ^{2}\mathcal{H}=-\gamma \phi _{\mathbf{%
\downarrow }}^{\ast }\phi _{\mathbf{\uparrow }},  \label{field}
\end{equation}%
with $\phi _{\mathbf{\downarrow },\mathbf{\uparrow }}\equiv \sqrt{X_{0}}\psi
_{\mathbf{\downarrow },\mathbf{\uparrow }}$, $\tau \equiv t/t_{0}$, $x\equiv
X/X_{0}$, $\mathcal{H\equiv }H/h_{c}$ and $\beta \equiv \mathbf{m}_{\mathbf{%
\downarrow }\mathbf{\uparrow }}\cdot \mathbf{m}_{\mathbf{\uparrow }\mathbf{%
\downarrow }}t_{0}/(\hbar SX_{0}),\eta \equiv t_{0}\Delta /2$, $\gamma
\equiv \mathrm{N}\varepsilon _{0}\omega _{L}^{2}m_{\mathbf{\downarrow }%
\mathbf{\uparrow }}X_{0}/(h_{c}S)$, $\kappa \equiv X_{0}\omega _{L}/c$
measured in natural units of time and coordinate, $t_{0}=\hbar /(\mu _{0}m_{%
\mathbf{\downarrow }\mathbf{\uparrow }}h_{c})$,$\quad X_{0}=\sqrt{\hbar
t_{0}/m}$, where $h_{c}$ is a magnetic-field strength and $\mathrm{N}$ the
total atom number. The rescaled wave function is subject to normalization $%
N_{\mathbf{\downarrow }}+N_{\mathbf{\uparrow }}\equiv \int_{-\infty
}^{+\infty }\left[ |\phi _{\mathbf{\uparrow }}(x)|^{2}+|\phi _{\mathbf{%
\downarrow }}(x)|^{2}\right] dx=1.$ If collisions between atoms in different
spin states are taken into account, which may be controlled by the Feshbach
resonance \cite{4}, $\beta $ in Eq. (\ref{phi}) combines contributions from
the MLFE and direct interactions. On proper rescaling, the same system of
Eqs. (\ref{phi}),(\ref{field}) applies to a degenerate gas of fermions \cite%
{Fermi1,Fermi2} with spin $1/2$, in which $\psi _{\mathbf{\downarrow }}$ and 
$\psi _{\mathbf{\uparrow }}$ represent two spin components, coupled by
magnetic field $\mathcal{H}$, and asymmetry $\eta $ is imposed by dc
magnetic field.

Equation (\ref{field}) can be solved using the respective Green's function 
\cite{Bang}, $\mathcal{H}\left( x,t\right) =A\exp (i\kappa x)+C\exp
(-i\kappa x)-\left( \gamma /2\kappa \right) \int_{-\infty }^{+\infty }\sin
\left( \kappa \left\vert x-x^{\prime }\right\vert \right) \phi _{\mathbf{%
\downarrow }}^{\ast }\left( x^{\prime },t\right) \phi _{\mathbf{\uparrow }%
}\left( x^{\prime },t\right) dx^{\prime }$, where constants $A$ and $C$
represent the solution of the corresponding homogeneous equation, i.e., they
are amplitudes of two incident counterpropagating microwaves. Since the MW
wavelength is far larger than the condensate size \textbf{(}$\kappa \sim
10^{-5}$\textbf{)}, one may set $\exp (i\kappa x)\approx 1$ and $\sin \left(
\kappa \left\vert x-x^{\prime }\right\vert \right) /\kappa \approx
\left\vert x-x^{\prime }\right\vert $ in the domain occupied by the
condensate, to simplify the solution: $\mathcal{H}\left( x,t\right) =%
\mathcal{H}_{0}-\left( \gamma /2\right) \int_{-\infty }^{+\infty }\left\vert
x-x^{^{\prime }}\right\vert \phi _{\mathbf{\downarrow }}^{\ast }\left(
x^{\prime },t\right) \phi _{\mathbf{\uparrow }}\left( x^{\prime },t\right)
dx^{\prime },$ where $\mathcal{H}_{0}\equiv A+C$ is made real by means of a
phase shift. The form of the magnetic field at $|x|\rightarrow \infty $
resembles that of a one-dimensional \textit{artificial magnetic monopole} 
\cite{monopole}: $\mathcal{H}_{\mathrm{asympt}}\left( x\right) =-\gamma
N_{\Updownarrow }|x|,~N_{\Updownarrow }\equiv \int_{-\infty }^{+\infty }\phi
_{\mathbf{\downarrow }}^{\ast }\left( x\right) \phi _{\mathbf{\uparrow }%
}\left( x\right) dx.$

Substituting the solution for $\mathcal{H}$ in Eq. (\ref{phi}), we arrive at
the final form of the GPEs:%
\begin{gather}
i\frac{\partial \phi _{\mathbf{\downarrow }}}{\partial \tau }=-\frac{1}{2}%
\frac{\partial ^{2}\phi _{\mathbf{\downarrow }}}{\partial x^{2}}-\mathcal{H}%
_{0}\phi _{\mathbf{\uparrow }}+\eta \phi _{\mathbf{\downarrow }}-\beta
\left\vert \phi _{\mathbf{\uparrow }}\right\vert ^{2}\phi _{\mathbf{%
\downarrow }}  \notag \\
+\frac{\gamma }{2}\phi _{\mathbf{\uparrow }}(x)\int_{-\infty }^{+\infty
}\left\vert x-x^{^{\prime }}\right\vert \phi _{\mathbf{\downarrow }}\left(
x^{\prime },t\right) \phi _{\mathbf{\uparrow }}^{\ast }\left( x^{\prime
},t\right) dx^{\prime },  \label{phia}
\end{gather}%
\begin{gather}
i\frac{\partial \phi _{\mathbf{\uparrow }}}{\partial \tau }=-\frac{1}{2}%
\frac{\partial ^{2}\phi _{\mathbf{\uparrow }}}{\partial x^{2}}-\mathcal{H}%
_{0}\phi _{\mathbf{\downarrow }}-\eta \phi _{\mathbf{\uparrow }}-\beta
\left\vert \phi _{\mathbf{\downarrow }}\right\vert ^{2}\phi _{\mathbf{%
\uparrow }}  \notag \\
+\frac{\gamma }{2}\phi _{\mathbf{\downarrow }}(x)\int_{-\infty }^{+\infty
}\left\vert x-x^{\prime }\right\vert \phi _{\mathbf{\downarrow }}^{\ast
}\left( x^{\prime },t\right) \phi _{\mathbf{\uparrow }}\left( x^{\prime
},t\right) dx^{\prime }.  \label{phib}
\end{gather}%
Thus, the MLFE gives rise to two nonlinear terms: the one $\sim \beta $
accounts for short-range interaction, while the integral term represents the
long-range interaction, which is \emph{locally repulsive}, but \emph{%
globally attractive} because the repulsion kernel, $\left\vert x-x^{\prime
}\right\vert $, growing at $|x|\rightarrow \infty $,\ suggests a possibility
of self-trapping. The mechanism of creating bright solitons by the spatially
growing strength of local self-repulsion was proposed in Ref. \cite%
{Barcelona}, and then extended for nonlocal %optical \cite{Yingji} and
dipolar-BEC \cite{Raymond} settings; however, that mechanism was imposed by
appropriately engineered spatial modulation of the nonlinearity, while here
we consider the self-trapping in free space.

%Along with norm (\ref{N}) and momentum, $P=i\sum_{\mathbf{\downarrow },%
%$\mathbf{\uparrow }}\int_{-\infty }^{+\infty }\phi _{\mathbf{\downarrow },%
%\mathbf{\uparrow }}\partial _{x}\phi _{\mathbf{\downarrow },\mathbf{\uparrow
%}}^{\ast }dx$, Eqs. (\ref{phia}) and (\ref{phib}) conserve the energy
%(Hamiltonian),%
%\begin{gather}
%E=\int_{-\infty }^{+\infty }\left[ \frac{1}{2}\left( \left\vert \partial
%_{x}\phi _{\mathbf{\downarrow }}\right\vert ^{2}+\left\vert \partial
%_{x}\phi _{\mathbf{\uparrow }}\right\vert ^{2}\right) -\left( \mathcal{H}%
%_{0}^{\ast }\phi _{\mathbf{\downarrow }}^{\ast }\phi _{\mathbf{\uparrow }}+%
%\mathcal{H}_{0}\phi _{\mathbf{\downarrow }}\phi _{\mathbf{\uparrow }}^{\ast
%}\right) \right.  \notag \\
%\left. -\beta \left\vert \phi _{\mathbf{\downarrow }}\right\vert
%^{2}\left\vert \phi _{\mathbf{\uparrow }}\right\vert ^{2}\right] dx+\frac{%
%\gamma }{2}\int \int \phi _{\mathbf{\downarrow }}^{\ast }(x)\phi _{\mathbf{%
%\uparrow }}(x)\phi _{\mathbf{\downarrow }}^{\ast }(x^{\prime })\phi _{%
%\mathbf{\uparrow }}(x^{\prime })\left\vert x-x^{\prime }\right\vert
%dxdx^{\prime }.  \label{E}
%\end{gather}

The symmetric version of Eqs. (\ref{phia}) and (\ref{phib}), with $\eta =0$,
may be combined into separate equations for $\phi _{\pm }\equiv \phi
_{\downarrow }\pm \phi _{\uparrow }$, with trapping (for $+$) and expulsive
(for $-$) potentials, respectively. Therefore, this system has only
symmetric solutions ($\phi _{-}=0$). At $\left\vert x\right\vert \rightarrow
\infty $, Eqs. (\ref{phia}) and (\ref{phib}) take the linear asymptotic
form, $i\partial _{\tau }\phi _{\downarrow \uparrow }=-(1/2)\partial
_{xx}^{2}\phi _{\downarrow \uparrow }+\left( \gamma N_{\Updownarrow
}/2\right) |x|\phi _{\uparrow \downarrow }$, hence solutions for the
asymmetric system ($\eta \neq 0$) have symmetric asymptotic tails too, $\phi
_{\downarrow }=\phi _{\uparrow }\sim \exp \left( -(2/3)\sqrt{\gamma
N_{\Updownarrow }}|x|^{3/2}\right) $.

\begin{figure}[tbp]
\begin{center}
\includegraphics{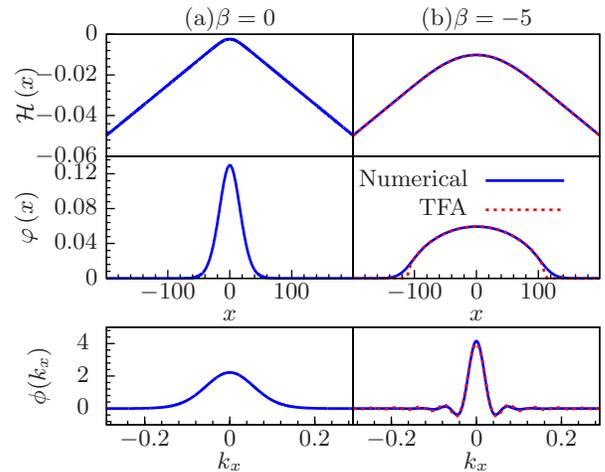}
\end{center}
\caption{(Color online) (a) The GS wave function, $\protect\varphi (x)$,
along with its Fourier transform, $\protect\phi \left( k_{x}\right) $, and
magnetic field, $\mathcal{H}\left( x\right) $, for $\protect\beta =0$, with $%
\tilde{\protect\mu}=4.26$. (b) Comparison of the GS, as predicted by the TFA
[Eqs. (\protect\ref{^2}), (\protect\ref{H(x)}) and (\protect\ref{TFF})]
(dashed curves) and found numerically (solid lines) for $\protect\beta =-5$\
with $\protect\mu =0.079$. In both plots, $\protect\eta =0$, $H_{0}=0$, and $%
\protect\gamma =10^{-3}$.}
\label{fig1}
\end{figure}

The GS of system (\ref{phia}), (\ref{phib}) with $\eta =0$ and chemical
potential $\mu $ is sought for as $\phi _{\mathbf{\downarrow \uparrow }%
}=e^{-i\mu t}\varphi (x)$, where real $\varphi $\ satisfies equation 
\begin{equation}
\tilde{\mu}\varphi =\left[ -\frac{1}{2}\frac{d^{2}}{dx^{2}}-\beta \varphi
^{2}+\frac{\gamma }{2}\int_{-\infty }^{+\infty }\left\vert x-x^{\prime
}\right\vert \varphi ^{2}\left( x^{\prime }\right) dx^{\prime }\right]
\varphi ,  \label{eff}
\end{equation}%
with $\mu \equiv \tilde{\mu}-\mathcal{H}_{0}$. Thus, $\mathcal{H}_{0}$ only
shifts the chemical potential in the zero-detuning system. For $\beta =0$,
it follows from Eq. (\ref{eff}) that $\tilde{\mu}$\ and the magnetic field
obey scaling relations: $\left\{ \tilde{\mu}(\gamma ),\mathcal{H}\left(
x;\gamma \right) -\mathcal{H}_{0}\right\} =\left( \gamma /\gamma _{0}\right)
^{2/3}\left\{ \tilde{\mu}\left( \gamma =\gamma _{0}\right) ,\mathcal{H}%
\left( x;\gamma =\gamma _{0}\right) -\mathcal{H}_{0}\right\} $, where $%
\gamma _{0}$ is a fixed constant. Thus, for $\eta =0$ and $\beta =0$, all
the GSs may be represented by a single one, plotted in Fig. \ref{fig1}(a),
which was found by means of the imaginary-time method %\cite{IT}
. This is a hybrid soliton, built of the self-trapped matter wave coupled to
the subwavelength deformation of the magnetic field. 

In the presence of the local self-repulsion ($\beta <0)$, the GS can be
found with the help of the Thomas-Fermi approximation (TFA), which neglects
the second derivative in Eq. (\ref{eff}):% \cite{6}
\begin{equation}
\varphi _{\mathrm{TFA}}^{2}(x)=\left\{ 
\begin{array}{c}
(1/4)\sqrt{\gamma /|\beta |}\cos \xi ,~\mathrm{at}~|\xi |<\pi /2, \\ 
0,~\mathrm{at}~~|\xi |>\pi /2,%
\end{array}%
\right.  \label{^2}
\end{equation}

\begin{equation}
\mathcal{H}_{\mathrm{TFA}}\left( x\right) +\mu =\left\{ 
\begin{array}{c}
(1/4)\sqrt{\gamma |\beta |}\cos \xi ,~\mathrm{at}~~|\xi |<\pi /2, \\ 
-\sqrt{\gamma |\beta |}\left( |\xi |-\pi /2\right) /4,~\mathrm{at}~~|\xi
|>\pi /2,%
\end{array}%
\right.  \label{H(x)}
\end{equation}%
where $\xi \equiv \sqrt{\gamma /|\beta |}x$. An example, displayed in Fig. %
\ref{fig1}(b), shows very good agreement of the TFA with the numerical
solution. Thus, the globally attractive long-range interaction induced by
the MLFE creates the self-trapped GS, overcoming the arbitrarily strong
self-repulsive contact interaction.

The time-of-flight spectrum produced by releasing the condensate can be used
in the experiment to detect the solitons predicted here, characterized by
their distribution over the longitudinal momentum,$\ \phi \left(
k_{x}\right) \equiv \int_{-\infty }^{+\infty }e^{-ik_{x}x}\varphi (x)dx$, as
shown in Fig. \ref{fig1} for $\beta =0$ and $-5$. In particular, the TFA
yields%
\begin{equation}
\phi _{\mathrm{TFA}}\left( k_{x}\right) =\frac{\pi \left( |\beta |/\gamma
\right) ^{1/4}}{8\Gamma \left( 5/4+\sqrt{|\beta |/4\gamma }k_{x}\right)
\Gamma \left( 5/4-\sqrt{|\beta |/4\gamma }k_{x}\right) },  \label{TFF}
\end{equation}%
where $\Gamma $\ is the Gamma-function, see the right bottom panel in Fig. %
\ref{fig1}. Note that expression (\ref{TFF}) vanishes at $k_{x}=\pm 2\sqrt{%
\gamma /|\beta |}\left( 5/4+n\right) ,~n=0,1,2,...$\ . The strong
compression of the soliton in the momentum space, evident in Fig. \ref{fig1}%
(b), may be used for the design of matter-wave lasers \cite{laser}, as the
released beam will feature high velocity coherence.

The existence of bright solitons in free space is related to the
modulational instability (MI) of flat states \cite{Berge}. The flat solution
to Eq. (\ref{eff}) is $\phi =\Phi _{0}\exp \left( -i\mu _{0}t\right) ,$ with
divergence of $\mu _{0}$ regularized by temporarily replacing $\left\vert
x-x^{\prime }\right\vert $ in Eq. (\ref{eff}) by $\left\vert x-x^{\prime
}\right\vert \exp \left( -\epsilon \left\vert x-x^{\prime }\right\vert
\right) $ with small $\epsilon >0$. The MI analysis for small perturbations
with wavenumber $k$ and MI gain $\lambda $ proceeds, as usual, by
substituting $\phi =\Phi \left( x,t\right) \exp \left( i\chi \left(
x,t\right) \right) $, with $\left\{ \Phi ;\chi \right\} =\left\{ \Phi
_{0};-\mu _{0}t\right\} +\left\{ \Phi _{1}^{(0)};\chi _{1}^{(0)}\right\}
\exp \left( ikx+\lambda t\right) $, where $\left\{ \Phi _{1}^{(0)};\chi
_{1}^{(0)}\right\} $ are perturbation amplitudes. The subsequent
linearization and then setting $\epsilon =0$ yields $\lambda ^{2}=-\left(
k^{4}/4-\beta \Phi _{0}^{2}k^{2}-\gamma \Phi _{0}^{2}\right) .$ Due to the
nonlocality, $\lambda ^{2}$ does not vanish at $k^{2}\rightarrow 0$, in
contrast with local models \cite{Berge}. The MI is \emph{always present}, as 
$\lambda ^{2}$ remains positive at $k^{2}<2\left\vert \Phi _{0}\right\vert
\left( \sqrt{\beta ^{2}\Phi _{0}^{2}+\gamma }+\beta \left\vert \Phi
_{0}\right\vert \right) $. Thus, arbitrarily strong local self-repulsion,
with $\beta <0$, does not suppress the MI.

\begin{figure}[tbp]
\begin{center}
\includegraphics{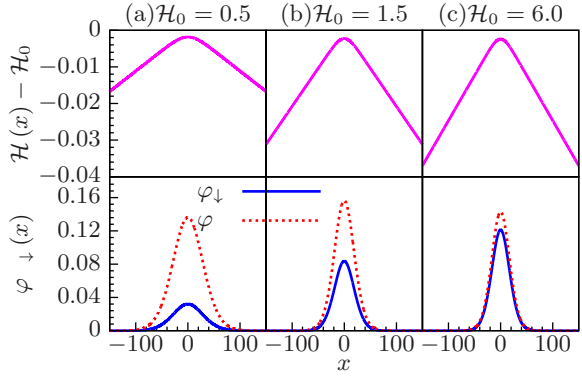}
\end{center}
\caption{(Color online) Numerically found profiles of the GS wave functions
and magnetic field in the system with $\protect\eta =1$, $\protect\beta =0$,
and $\protect\gamma =10^{-3}$, for $\mathcal{H}_{0}=0.5$ (a), $1.5$ (b), $%
6.0 $ (c). The respective chemical potentials are $\protect\mu =-1.12$ (a), $%
-1.80$ (b), $-6.079$ (c). For the strongly asymmetric soliton in (a), the
analytical approximation predicts the amplitude ratio of the two components $%
0.250$, while its numerically found counterpart is $0.235$.}
\label{fig2}
\end{figure}

\begin{figure}[tbp]
\begin{center}
\includegraphics{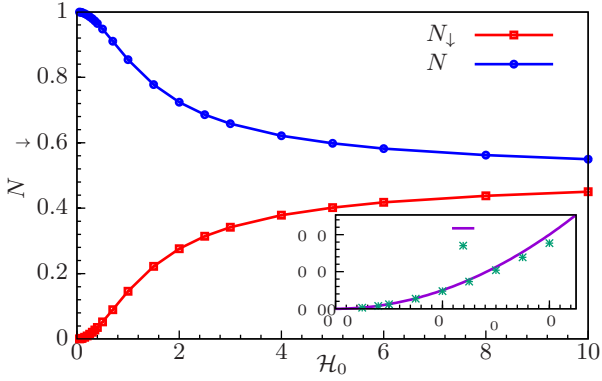}
\end{center}
\caption{(Color online) The relative share of the total norm of each
component in the system with $\protect\gamma =10^{-3}$, $\protect\beta =0$, $%
\protect\eta =1$, vs. the background magnetic field, $\mathcal{H}_{0}$. The
inset shows the dependence for $N_{\downarrow }$ at small values of $%
\mathcal{H}_{0}$ (stars) vis-a-vis the analytical prediction.}
\label{fig3}
\end{figure}

In the system with detuning, i.e., $\eta \neq 0$ in Eqs. (\ref{phia}) and (%
\ref{phib}), the background magnetic field $\mathcal{H}_{0}$ is an essential
parameter. Figure \ref{fig2} plots the GS wave functions, $\varphi _{\mathbf{%
\downarrow \uparrow }}\left( x\right) $, and the corresponding magnetic
field, $\mathcal{H}\left( x\right) $, for different values of $\mathcal{H}%
_{0}$. The GS exhibits asymmetry between the lower- and higher-energy
components at $\mathcal{H}_{0}\lesssim \eta $, while large $\mathcal{H}_{0}$
suppresses the asymmetry, as seen in Fig. \ref{fig3}, which displays the
scaled norms of the two components, $N_{\mathbf{\downarrow \uparrow }}$,
versus $\mathcal{H}_{0}$. 

For $\eta \gg \mathcal{H}_{0}$, strongly asymmetric GSs can be found using
the stationary version of Eqs. (\ref{phia}) and (\ref{phib}) with chemical
potential $\mu =-\eta +\Delta \mu $, $\left\vert \Delta \mu \right\vert \ll
\eta $. Then, Eq. (\ref{phia}) eliminates the weak component in favor of the
strong one: $\varphi _{\mathbf{\downarrow }}\approx \left( 2\eta \right)
^{-1}\mathcal{H}_{0}\varphi _{\mathbf{\uparrow }}$, and$~N_{\downarrow
}\approx \mathcal{H}_{0}^{2}/\left( 4\eta ^{2}\right) $, which agrees well
with numerical results, as shown by the inset in Fig. \ref{fig3}. The
substitution of this into Eq. (\ref{phib}) yields%
\begin{gather}
\left( \Delta \mu +\frac{\mathcal{H}_{0}^{2}}{2\eta }\right) \varphi _{%
\mathbf{\uparrow }}=-\frac{1}{2}\frac{d^{2}\phi _{\mathbf{\uparrow }}}{dx^{2}%
}-\frac{\beta \mathcal{H}_{0}^{2}}{4\eta ^{2}}\varphi _{\mathbf{\uparrow }%
}^{3}  \notag \\
+\frac{\gamma \mathcal{H}_{0}^{2}}{8\eta ^{2}}\varphi _{\mathbf{\uparrow }%
}(x)\int_{-\infty }^{+\infty }\left\vert x-x^{\prime }\right\vert \varphi _{%
\mathbf{\uparrow }}^{2}(x^{\prime })dx^{\prime },  \label{Delta}
\end{gather}%
which is actually tantamount to Eq. (\ref{eff}). The respective small
deformation of the magnetic field is $\mathcal{H}\left( x\right) =\mathcal{H}%
_{0}-\left( \gamma \mathcal{H}_{0}/4\eta \right) \int_{-\infty }^{+\infty
}\phi _{\mathbf{\uparrow }}^{2}\left( x^{\prime }\right) \left\vert
x-x^{\prime }\right\vert dx^{\prime }$.

We have confirmed the stability of all the GS states by direct simulations
of Eqs. (\ref{phia}) and (\ref{phib}) with randomly perturbed initial
conditions. For the symmetric system with $\eta =\beta =0$, the
above-mentioned scaling implies that the stability of a single GS guarantees
the stability of all GSs, while for the detuned system the stability had to
be checked by varying $\mathcal{H}_{0}$ at fixed $\gamma $ and $\eta $.
Further, the stability for $\eta =\beta =0$ is predicted by the\textit{\
anti-Vakhitov-Kolokolov} (anti-VK) criterion, which states that the
necessary stability condition for \emph{bright}\textit{\ }solitons supported
by the \emph{repulsive }nonlinearity is $d\mu /dN>0$ \cite{Hidetsugu} (the
VK criterion proper, which pertains to attractive nonlinearity, is $d\mu
/dN<0$ \cite{VK,Berge}). Although $N=1$ was fixed above, the criterion can
be applied by means of rescaling which fixes $\gamma $ and liberates $N$. As
a result, the scaling relation $\tilde{\mu}\sim \gamma ^{2/3}$ is replaced
by $\tilde{\mu}\sim $ $N^{2/3}$, hence the criterion holds.

Summarizing, we have explored the MLFE (magnetic local-field effect) in the
BEC built of two atomic states coupled by the MW (microwave) field. We have
deduced the system of evolution equations for the matter-wave components and
MW magnetic field, which demonstrate that the subwavelength distortion of
the magnetic field by perturbations of the local atom density induces short-
and long-range interactions between the BEC components. The same equations
apply to the spinor wave function of a fermionic gas coupled to the MW
magnetic field. The model produces the self-trapped GS (ground state) in the
form of the hybridized BEC-MW subwavelength solitons, which may be
considered as counterparts of hybrid exciton-polariton solitons in the
dissipation-free system. Basic characteristics of the solitons were obtained
analytically. The release of the solitons from the cigar-shaped trap \ may
be used as a source of coherent matter waves for an atom laser. The flat
states in the present \ system are subject to the modulational instability,
which is naturally related to the existence of the bright solitons.

It is straightforward to extend the analysis for molecular BECs, with the
transition between two rotational states driven by the electric MW field.
Such ultracold molecular gases have a potential for quantum simulations of
condensed-matter physics \cite{zoller}. An interesting extension may be the
analysis of the system with a \textit{three-component} bosonic wave function
corresponding to spin $F=1$ \cite{spinor1}, in which a single MW field
couples components with $m_{F}=\pm 1$\ to the one with $m_{F}=0$. Other
relevant directions for the extension are search for excited states in the
system, in addition to the GS, and the analysis of the two-dimensional
setting. Furthermore, it is relevant to investigate a potential effect of
the MLFE on quantum precision measurements.

%\begin{acknowledgement}
G.D. acknowledges the support of the National Basic Research Program of
China ("973" Program, No. 2011CB921602), the National Natural Science
Foundation of China (No. 11034002), and the Research Fund for the Doctoral
Program of Higher Education of China (No. 20120076110010). The work of
B.A.M. was supported, in a part, by grant No. B12024 from the Program of
Introducing Talents of Discipline to Universities (China). 
%\end{acknowledgement}

\end{document}